# Impact of Label Noise from Large Language Models Generated Annotations on Evaluation of Diagnostic Model Performance

Mohammadreza Chavoshi[1], Hari Trivedi[1], Janice Newsome[1], Aawez Mansuri[1], Chiratidzo Rudado Sanyika[1], Rohan Satya Isaac[1], Frank Li[1], Theo Dapamede[1], Judy Gichoya[1]

[1]Department of Radiology, Emory University, Atlanta, GA, USA

## Abstract

**Background:**
Large language models (LLMs) are increasingly used to generate labels from radiology reports for large-scale AI evaluation. However, errors in LLM-generated labels may bias performance estimation, especially under varying disease prevalence and model quality. This study quantifies the impact of LLM label noise on downstream diagnostic model evaluation.

**Methods:**
We developed a simulation framework to evaluate how LLM label errors affect observed model performance. A synthetic dataset of 10,000 cases was generated across varying disease prevalence levels (10%, 30%, 70%, 90%). LLM sensitivity and specificity were varied independently from 90% to 100%. We simulated models with true performance ranging from 90% to 100% sensitivity and specificity. Observed performance was calculated using LLM-generated labels as the reference standard. Best- and worst-case performance bounds were calculated analytically, and empirical uncertainty distributions were obtained via 5,000 Monte Carlo trials per condition.

**Results:**
Observed performance calculation was highly sensitive to LLM label quality, with estimation bias strongly modulated by disease prevalence. In low-prevalence settings, small reductions in LLM specificity substantially underestimated sensitivity. At 10% prevalence, an LLM with 95% specificity yielded an observed sensitivity of ~53% despite a perfect model. In high-prevalence conditions, reduced LLM sensitivity led to an underestimation of specificity. Monte Carlo simulations revealed consistent downward bias, with estimated values often falling below the true model performance even when within theoretical error bounds.

**Conclusion:**

LLM-generated labels can introduce systematic, prevalence-dependent bias into model evaluation. Specificity is more critical in low-prevalence tasks, while sensitivity is more impactful in high-prevalence settings. These findings underscore the need for prevalence-aware prompt engineering and error reporting when using LLMs for post-deployment model evaluation in clinical AI applications.

# Introduction

The performance of medical imaging AI models must be validated beyond controlled research settings to ensure their generalizability to real-world clinical practice. In recent years, there has been a growing trend toward large-scale, post-deployment evaluations of AI tools in medical imaging to assess their true clinical utility and robustness(1). A key obstacle in conducting such evaluations is the need for reliable ground-truth labels across vast imaging datasets. In clinical radiology, the free-text radiology report associated with each imaging study is often used as a reference standard. This approach is far more scalable than manual image annotation, which is time-consuming and labor-intensive, particularly when expert review is required at scale.

To address this, a wide range of natural language processing (NLP) techniques have been developed to extract structured labels from unstructured radiology reports. These range from simple keyword-based systems to advanced deep learning models. Rule-based methods like CheXpert(2) and cTAKES(3) use medical lexicons to manage common variations and expressions of uncertainty, while more recent transformer-based approaches (e.g., BERT, RadReportAnnotator) learn from labeled data to interpret complex language patterns (4,5). Many contemporary systems combine both strategies in hybrid pipelines(6). Despite this progress, achieving human-level labeling accuracy remains difficult due to the inherent diversity and nuance of clinical language(7).

Large language models (LLMs) offer a promising new approach for extracting structured labels from radiology reports using prompt-based learning, without requiring task-specific retraining. By leveraging general language understanding and contextual reasoning, LLMs can interpret clinical text when guided by well-designed prompts. For example, a prompt-engineered version of GPT-4 was able to extract oncologic findings from CT reports with 95–98% agreement with expert annotations, underscoring the potential of LLMs as efficient and adaptable alternatives to other NLP methods(8).

However, when the labels extracted by LLMs are used as a reference standard, any errors in those labels can propagate and distort the evaluation of downstream AI model performance(9). This can mislead assessments of a model's effectiveness, especially in large-scale studies. As a result, there is increasing focus on prompt engineering, tailoring instructions, formats, or examples to improve the accuracy of LLM-extracted labels(10). This process involves a trade-off: broad prompts may increase sensitivity but lead to more false positives, while narrow prompts may boost precision at the cost of missing true findings. Studies suggest that overly specific prompts can reduce accuracy by misguiding the model, whereas context-rich or few-shot prompts often improve performance(11). Optimizing this balance is critical for ensuring LLMs produce reliable reference labels for clinical applications.

Given these challenges, we designed a simulated study to systematically examine how LLM-based labeling errors impact the real-world evaluation of AI models. Specifically, we aimed to answer the following three questions:

1. How do errors in LLM labeling affect observed model performance metrics across different disease prevalence scenarios?
2. How does improving LLM sensitivity and specificity influence the accuracy of performance estimation?
3. How does the inherent performance of the evaluated AI model influence the error introduced by LLM-derived labels?

## Method

### 1. Simulation Framework

All experiments were conducted using a unified simulation framework designed to evaluate how imperfections in LLM-generated labels affect the apparent performance of diagnostic models. The same definitions and procedures were used across all phases.

### 1.a. Sample Size

Each simulation used a synthetic dataset of 10,000 cases. Ground-truth disease status was generated by assigning a fixed number of positive and negative cases based on the specified prevalence.

### 1.b. LLM Annotation Simulation

The LLM was simulated as an imperfect labeler, with its sensitivity and specificity independently varied between 90% and 100% in 1% increments. For each pair of sensitivity and specificity values, LLM-generated labels were assigned by probabilistically applying these values to the ground-truth disease status.

*1.c. Confusion Matrix Construction*

For both the model and the simulated LLM, confusion matrix components were derived as follows:

• True Positives (TP) = sensitivity × number of true positives
• False Negatives (FN) = number of true positives − TP
• True Negatives (TN) = specificity × number of true negatives
• False Positives (FP) = number of true negatives − TN

These components were used to generate predicted labels independently for the model and the LLM, according to their defined characteristics.

*1.d. Observed Model Performance Using LLM Annotations*

The LLM-derived labels were treated as the reference standard for evaluating model performance.

- Estimated (or observed?) Sensitivity was defined as the proportion of LLM-labeled positives that the model also predicted as positive.

$$\text{Estimated Sensitivity} = \frac{\text{TP}_{\text{model}} \cap \text{TP}_{\text{LLM}} + \text{FP}_{\text{model}} \cap \text{FP}_{\text{LLM}}}{\text{TP}_{\text{LLM}} + \text{FP}_{\text{LLM}}}$$

- Estimated Specificity was defined as the proportion of LLM-labeled negatives that the model also predicted as negative.

$$\text{Estimated Specificity} = \frac{\text{TN}_{\text{model}} \cap \text{TN}_{\text{LLM}} + \text{FN}_{\text{model}} \cap \text{FN}_{\text{LLM}}}{\text{TN}_{\text{LLM}} + \text{FN}_{\text{LLM}}}$$

*1.e. Calculate the Label-induced Error Range*

To account for the variability introduced by label noise, we calculated both best-case and worst-case estimates of estimated sensitivity and specificity for each simulated condition. These bounds reflect the plausible range of apparent model performance that could arise when evaluation is conducted using an imperfect reference standard.

- In the best-case scenario, agreement between the model and LLM is assumed to be maximized within the limits of their respective predictions. Specifically, model-labeled positives and negatives are aligned with LLM-labeled positives and negatives in a way that yields the greatest possible overlap. Any disagreement is minimized and assigned in a manner that favors concordance (e.g., excess model positives are assumed to fall within LLM-labeled positives wherever possible) (Figure 1a).
- In the worst-case scenario, the model and LLM are assumed to disagree as much as possible. Their predictions are arranged to minimize overlap, so the model appears to perform as poorly as the label mismatch allows. (Figure 1b).

Error range was defined as the difference between the best- and worst-case estimates for each performance metric, reflecting the range of observed values that could arise solely from misalignment between model predictions and imperfect LLM-derived annotations.

*1.f. Monte Carlo Simulation of Observed Performance*

To complement theoretical bounds, we used Monte Carlo simulation to estimate the empirical uncertainty in observed sensitivity and specificity under each LLM configuration (12). For each combination of LLM sensitivity and specificity, we conducted 5,000 independent trials. In each trial, model and LLM predictions were generated independently based on the ground truth, and model performance was evaluated using the LLM-generated labels as the reference standard. The maximum and minimum of the 5,000 observed values were recorded for each configuration.

*2. Evaluation of Disease Prevalence Effect*

To isolate the effect of disease prevalence, the model's true performance was fixed at 100% sensitivity and specificity. We evaluated disease prevalence at 10%, 30%, 70%, and 90%, representing both low- and high-prevalence conditions. Estimated performance metrics were computed at each prevalence level using LLM-derived labels as the reference standard. Heatmaps were generated to visualize observed performance across the full range of LLM sensitivities and specificities at selected prevalence levels.

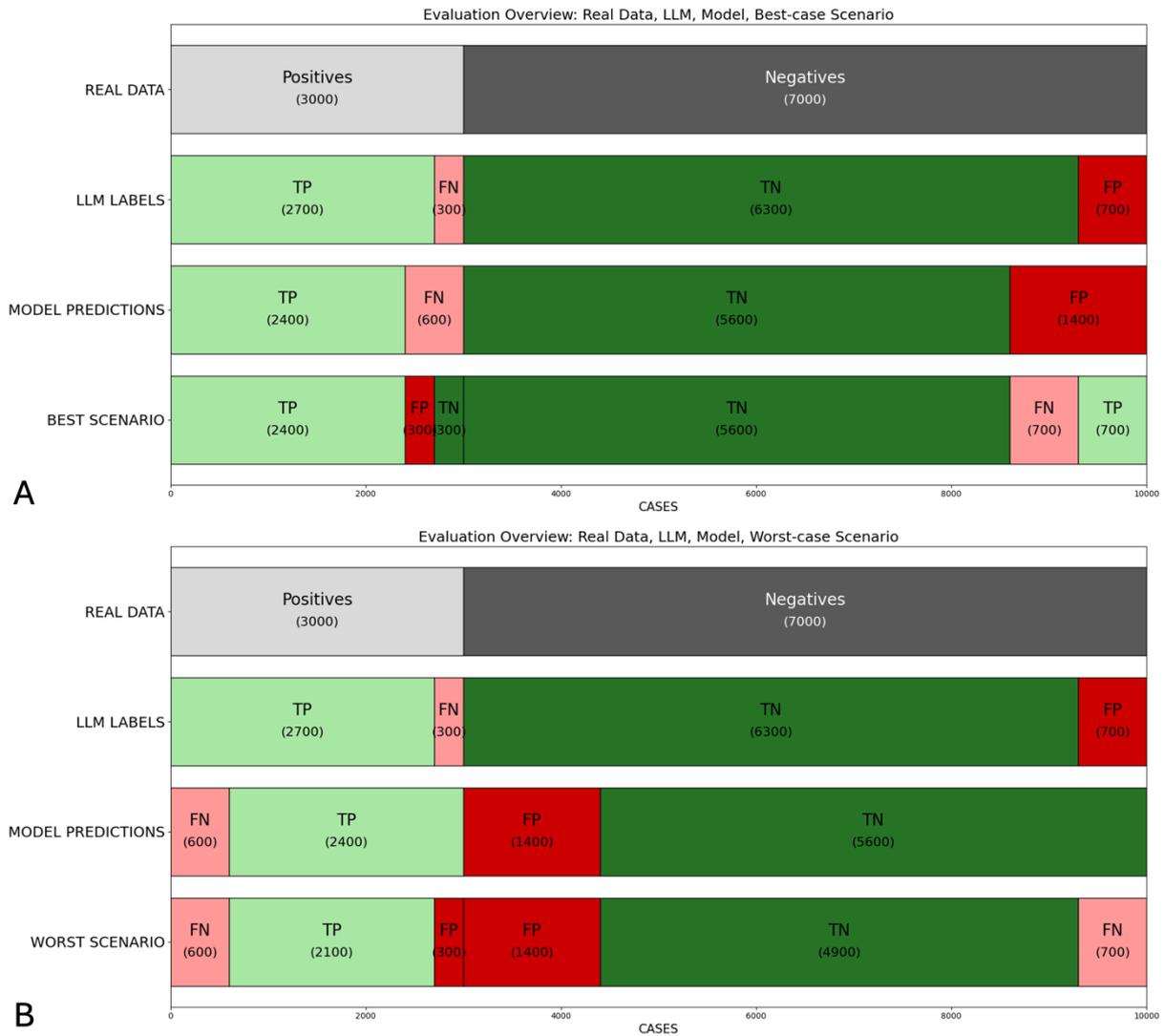

Figure 1. Visual overview of the simulated real data distribution, LLM-generated labels, model predictions, and the derived best-case and worst-case evaluation scenario. Here we simulated a 10000 population for a disease prevalence of 30%, LLM sensitivity and specificity of 90%, and model sensitivity and specificity of 80%. A) Best-case scenario for observed model performance. In this setting, the overlap between the model and LLM predictions is maximized, assuming the highest possible agreement. This results in the most favorable observed sensitivity (81.6%) and specificity (95.2%) when using LLM labels as the reference standard. B) Worst-case scenario for observed model performance. In this setting, disagreement between model and LLM predictions is maximized, assuming minimal overlap in the predicted cases. This yields the lowest observed sensitivity (61.8%) and specificity (74.2%) under the same labeling conditions.

## 3. Evaluation of Performance Estimation at Fixed Model Accuracy

To evaluate how LLM label quality affects performance estimation under realistic model conditions, we fixed model sensitivity and specificity at 95%. Prevalence was set at 10% and 30%, which are typical of many real-world clinical scenarios. LLM sensitivity and specificity were varied independently from 90% to 100% in 1% increments. For each configuration, estimated sensitivity and specificity were calculated, and best- and worst-case values were used to define the error range. These error ranges, along with empirical distributions from the Monte Carlo simulations, were visualized using forest plots to compare configurations and highlight the expected direction and magnitude of estimation error.

*4. Evaluation Across Varying Model Performance Levels*

To assess whether LLM labeling errors have a consistent impact across models with different accuracies, we simulated models with true performance set to 90% and 98% sensitivity and specificity, using a fixed disease prevalence of 10%. Estimated performance metrics were computed using the same procedure as above. Forest plots were used to visualize best-case, worst-case, and Monte Carlo error ranges, allowing direct comparison of label-induced estimation error across baseline model performance levels.

# Results

*1. Effect of Disease Prevalence on Observed Performance*

At a disease prevalence of 10%, observed sensitivity was highly vulnerable to errors in LLM specificity. As LLM specificity decreased from 100% to 90%, observed sensitivity dropped markedly, from 100% to approximately 53%, even when LLM sensitivity remained perfect. In contrast, observed specificity remained stable above 98%, even in the presence of moderate LLM noise. At 30% prevalence, this imbalance persisted but was attenuated. The sensitivity error range narrowed (79.4%–100%), indicating improved tolerance to LLM errors as prevalence increased. This highlights how rare disease conditions are more susceptible to label noise, which can severely bias sensitivity estimates downward (Figure 2).

In high-prevalence settings, the pattern is reversed. At 90% prevalence, observed specificity became more sensitive to errors in LLM sensitivity. Reducing LLM sensitivity from 100% to 90% caused observed specificity to fall from 100% to ~53%, even with perfect LLM specificity. Meanwhile, observed sensitivity remained consistently high across all conditions. (Supplementary Figure 1).

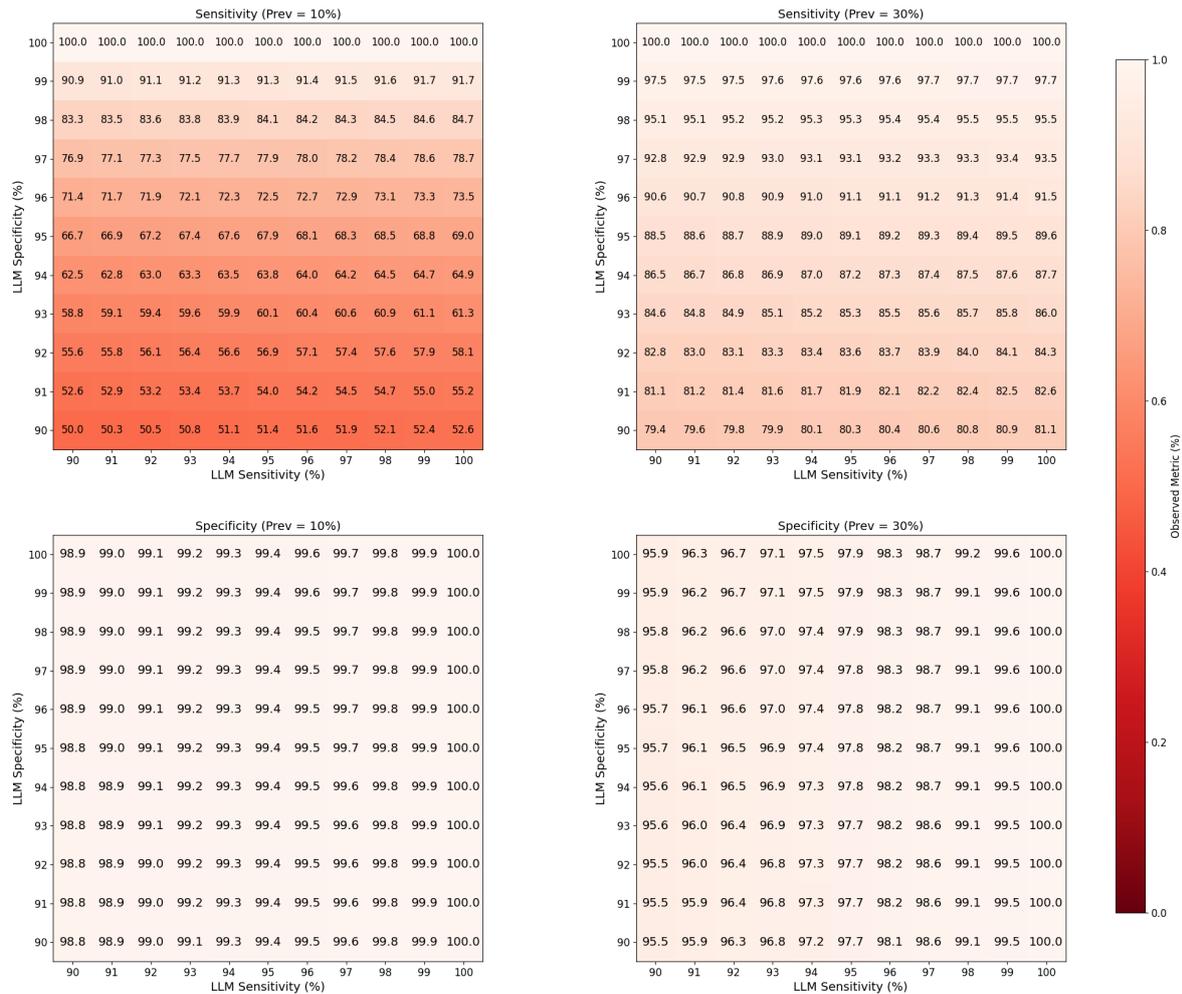

*Figure 2. Observed performance at disease prevalence 10% and 30%. Heatmaps showing observed sensitivity (top) and specificity (bottom) of a model with perfect performance (100% sensitivity and specificity) when evaluated against LLM-derived labels at 10% and 30% disease prevalence. Each grid cell corresponds to a different configuration of LLM sensitivity and specificity (ranging from 90% to 100%). At 10% prevalence, observed sensitivity is highly unstable with even small decreases in LLM specificity, while specificity remains relatively unaffected. At 30% prevalence, this pattern is still visible but less severe, with narrower uncertainty ranges. These results illustrate how false positives from label noise have a disproportionately large impact on observed sensitivity in low-prevalence settings.*

## 2. Error Range in Performance Estimation at Fixed Model Accuracy

Forest plots at 10% and 30% disease prevalence illustrate how uncertainty in estimated performance varies across all combinations of LLM sensitivity and specificity between 90% and 100% (Figure 3a). At 10% prevalence, the theoretical error in estimated sensitivity ranged from 0.00 (when both LLM sensitivity and specificity were 100%) to a maximum of 0.37, observed when LLM sensitivity was 90% and specificity was 95%. For estimated specificity, the error ranged from 0.00 to 0.06, with the highest uncertainty occurring when LLM specificity dropped below 95%.

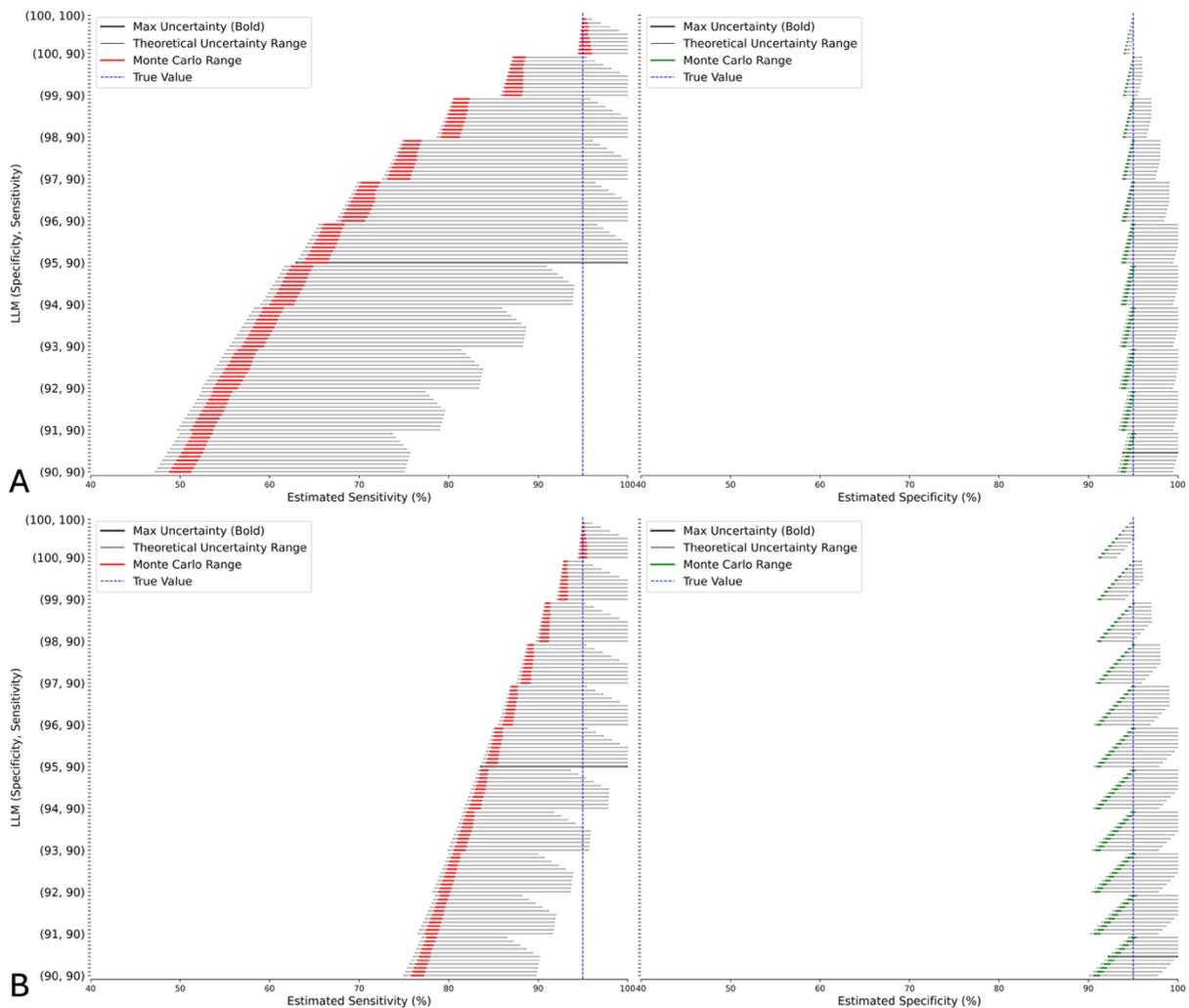

*Figure 3. A) Estimated performance at 10% disease prevalence. Forest plots showing theoretical and Monte Carlo uncertainty in estimated sensitivity (left) and specificity (right) across all combinations of LLM sensitivity and specificity between 90% and 100%, for a model with 95% true sensitivity and specificity at 10% disease prevalence. Each horizontal line represents the theoretical best- and worst-case performance estimates based on LLM labels; bold lines indicate configurations with the maximum uncertainty. The red and green bars represent the Monte Carlo range derived from 5,000 simulations. At this low prevalence, uncertainty in estimated sensitivity is substantial, particularly when LLM specificity is below 95%, and often fails to capture the model's true performance (blue dashed line). In contrast, specificity estimates are more stable, underscoring the asymmetric effect of label errors in low-prevalence conditions. B) Estimated performance at 30% disease prevalence. While label noise still introduces uncertainty, the sensitivity and specificity estimates are more concentrated around the true value, and Monte Carlo simulations frequently capture it within their range. Compared to 10% prevalence, both the theoretical and empirical uncertainty ranges are narrower, and the risk of systematic underestimation is reduced. These results suggest that the impact of LLM annotation quality is amplified at lower disease prevalence, where even small reductions in specificity can lead to substantial misestimation of model performance.*

Monte Carlo simulations validated these findings by showing that the range of observed sensitivity and specificity values across 5,000 trials which closely tracked the theoretical error bounds. Notably, in configurations where LLM specificity dropped below 95% at 10% prevalence—or below 93% at 30% prevalence—the entire theoretical and empirical distributions of estimated sensitivity fell below the model's true value of 95%. This indicates that under such labeling noise, accurate performance estimation becomes fundamentally unattainable.

Across all tested configurations, estimated sensitivity was consistently more vulnerable to labeling errors than specificity, particularly in low-prevalence settings. This was largely due to the disproportionate impact of false positives on sensitivity estimates when true positive cases are rare. Improving LLM sensitivity from 90% to 100% reduced uncertainty in sensitivity estimates by up to 6% and in specificity by about 1%. In contrast, increasing LLM specificity over the same range reduced uncertainty in sensitivity by as much as 24% and in specificity by 5%.

At 30% prevalence, performance estimates were notably more stable. The estimated sensitivity range narrowed to 79.4%–100%, and specificity remained above 95.5% across all LLM configurations. Moreover, the likelihood of underestimating true model performance—particularly in the Monte Carlo simulations, was substantially reduced compared to the 10% prevalence scenario (Figure 3b).

*3. Impact of Varying Model Performance*

Estimated sensitivity remained more affected by LLM labeling errors than specificity across models with both high (98%) and moderate (90%) performance (Figure 4). Although the absolute uncertainty range narrowed slightly with higher model performance, the overall structure of the error remained consistent. Configurations with lower LLM specificity consistently produced the widest error range in estimated sensitivity and frequently failed to capture the true model performance within either the analytical or Monte Carlo bounds. In contrast, estimated specificity remained relatively stable and was less impacted by variations in both model quality and LLM error.

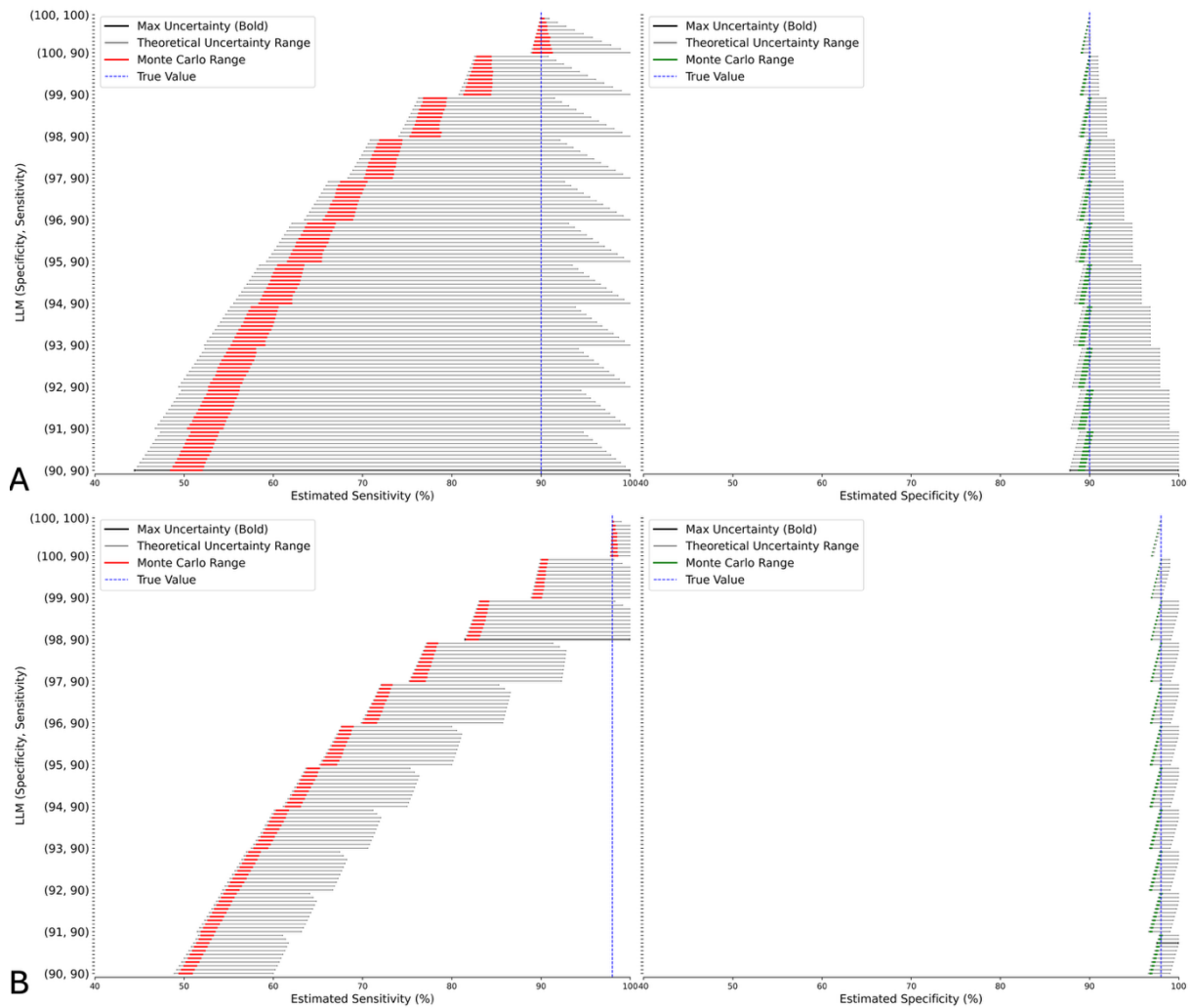

Figure 4. Estimated model performance across all LLM label quality configurations at 10% disease prevalence, for models with true performance of (A) 90% and (B) 98% sensitivity and specificity. Each row corresponds to a specific LLM sensitivity-specificity pair ranging from 90% to 100%. Gray lines denote theoretical uncertainty bounds, red and green bars represent the Monte Carlo simulation range, and dashed blue lines indicate the model's true performance. In both models, label noise from the LLM caused substantial uncertainty in estimated sensitivity. The magnitude of uncertainty decreased as the model's true performance improved (from 90% to 98%), but the pattern of sensitivity being more vulnerable to LLM errors in low-prevalence settings remained consistent.

## Discussion

NLP methods are increasingly used to annotate radiology datasets by extracting structured labels from free-text reports, as manual labelling is impractical at scale(13). However, consistent label extraction remains a significant challenge. Radiologists often use varied expressions, synonyms, and shorthand, introducing substantial variability in clinical language. This persists despite standardized vocabularies such as the Unified Medical Language System (UMLS)(14), Medical Subject Headings (MeSH)(15) and RadLex(16,17), which aim to unify radiology reporting terminology. In practice, however, many expressions used in reports fall outside these controlled vocabularies. One analysis found that nearly one-third of clinician-used terms were not included in RadLex(18). Moreover, radiology reports frequently contain nuanced or ambiguous phrasing, such as "possible pneumonia" or double negatives, that can be difficult for rule-based systems to parse. Recent work has shown that LLMs like GPT-4 can outperform traditional approaches on this task, achieving high F1 scores and increasing their appeal for large-scale annotation(19).

In this study, we systematically evaluated how errors in LLM-generated labels influence the estimation of model performance across a range of prevalence conditions and model qualities. Our simulations revealed consistent and interpretable patterns: observed performance metrics are highly sensitive to the type and magnitude of LLM annotation error, and these effects are strongly modulated by disease prevalence. In low-prevalence conditions, LLM specificity plays a pivotal role in preserving the accuracy of estimated sensitivity, while in high-prevalence settings, LLM sensitivity has a greater influence on estimated specificity. Across all scenarios, we observed that performance is often underestimated, even when the theoretical error bounds contain the true value, emphasizing the real-world risk of misjudging model quality due to imperfect LLM-derived labels.

These findings align with prior research showing that diagnostic performance metrics are not static and can vary with disease prevalence (20). In low-prevalence settings, common in radiology, even modest reductions in LLM specificity can produce more false positives than true positives, leading to systematic underestimation of sensitivity. For example, at 10% prevalence, an LLM with 95% specificity may generate a false positive rate that overwhelms the number of true positives, skewing observed sensitivity downward despite high overall LLM accuracy. Monte Carlo simulations confirmed this trend: when LLM specificity dropped below 99%, the full distribution of estimated sensitivities often fell below the model's true value. In contrast, specificity estimates remained relatively robust to LLM sensitivity errors. These results underscore that LLM specificity, not sensitivity, is the dominant factor driving label-induced bias in low-prevalence conditions. Accordingly, prompt engineering and quality control efforts should prioritize maximizing specificity when annotating rare findings which is common in the medical domain.

As prevalence increases, this relationship inverts: LLM sensitivity becomes more critical in avoiding false negatives that distort specificity estimates. This highlights the need for prevalence-aware prompt design, where prompt strategies are tuned based on the expected frequency of the label being extracted. Across both prevalence extremes, we observed that LLM label noise consistently introduced asymmetric and substantial estimation bias, even when the model itself had high accuracy.

Our Monte Carlo simulations added empirical insight to the theoretical bounds. While the best- and worst-case estimates captured the range of possible errors introduced by imperfect labeling, the simulations revealed a strong directional bias toward underestimation, particularly in low-prevalence conditions. The empirical distributions of estimated sensitivity were consistently skewed below the model's true performance, even when the theoretical range included it. Similar to prior diagnostic modeling work using Monte Carlo simulations to assess bias and variance (21,22), our results illustrate that under realistic conditions, LLM label noise does not just introduce uncertainty, it systematically depresses performance estimates. This reinforces the need to interpret LLM-derived metrics with caution, especially when applied to evaluate models on rare clinical findings.

Inaccurate or inconsistent labeling has been shown to significantly alter model behavior and reliability, even when model architecture and data remain constant (23). Our findings align with this, demonstrating that LLM labeling noise can produce substantial misestimation of performance. We observed that the structure of label-induced misestimation persisted across models with true sensitivity and specificity ranging from 90% to 98%. Although high-performing models exhibited narrower error ranges, the pattern of misestimation, driven by prevalence and LLM label quality, remained unchanged. Low LLM specificity led to underestimated sensitivity in rare conditions, while reduced LLM sensitivity affected specificity estimation in common conditions. These effects were most pronounced in lower-performing models but consistent in direction across all model qualities.

This study has several limitations. First, all simulations were based on synthetic datasets with predefined prevalence rates and fixed performance metrics for both the model and the LLM. While this controlled design allowed systematic exploration, it does not fully reflect the complexity of real-world data and LLM behaviour. Second, our analysis was limited to binary classification tasks, while many clinical applications involve multi-class or multi-label outputs that pose additional challenges for label extraction. Third, we assumed a perfect ground truth, whereas real-world datasets may contain noise, inter-observer variability, or uncertainty even in expert annotations. Additionally, LLM outputs can vary across repeated runs on the same input due to generation non-determinism. Parameters such as temperature and nucleus sampling (top-p) introduce randomness in token selection(24,25). meaning the same prompt may yield different outputs unless generation is carefully controlled. Likewise, deep learning models with stochastic components (e.g., dropout layers or ensemble variation) may generate non-reproducible predictions unless explicitly made deterministic(26). Our simulations did not model these additional sources of variability, potentially underestimating the full uncertainty present in real-world LLM-assisted evaluation pipelines. Finally, while Monte Carlo simulations helped characterize empirical uncertainty, they assume error independence and may not fully capture dependencies between LLM and model behavior in real clinical settings.

Taken together, our findings demonstrate that LLM-generated labels can introduce substantial and systematic bias into model performance evaluation, a bias that is strongly shaped by label prevalence. In low-prevalence conditions, imperfect LLM specificity inflates false positives and leads to significant underestimation of sensitivity. In high-prevalence settings, reduced LLM sensitivity increases false negatives, lowering observed specificity. These effects were consistent across models of varying quality and were confirmed empirically via Monte Carlo simulations, which revealed that underestimation is not just possible, but likely, under realistic labeling errors.

These insights have direct implications for prompt engineering and LLM deployment strategies. Prior studies have shown that prompt design can influence whether LLMs favor sensitivity or specificity in clinical tasks (27). Our results support a prevalence-aware prompting approach: when extracting rare findings, maximizing specificity should be prioritized to avoid overwhelming false positives; for common findings, optimizing sensitivity becomes more critical. Additionally, when LLM-generated labels are used to evaluate model performance, reporting the associated estimation error or uncertainty range can help downstream users interpret results more cautiously and avoid overly optimistic or pessimistic conclusions. As LLMs become further embedded in clinical evaluation pipelines, aligning prompt strategies with label prevalence—and disclosing expected error—will be essential for ensuring valid and trustworthy model assessment.

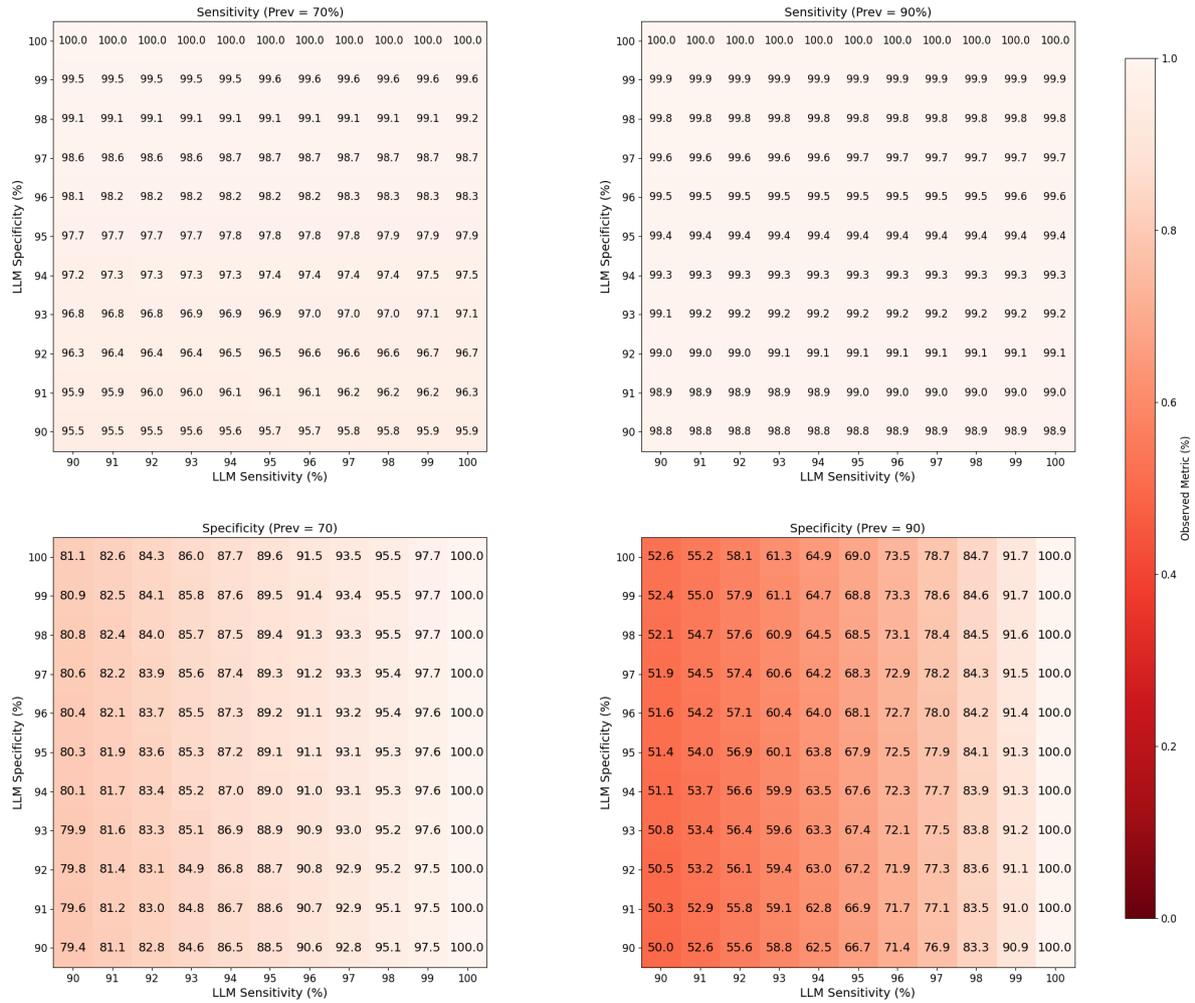

Supplementary Figure 1. Observed performance at disease prevalence 70% and 90%. Heatmaps showing observed sensitivity (top) and specificity (bottom). At high prevalence, the pattern is reversed: observed specificity becomes highly sensitive to LLM sensitivity errors, while observed sensitivity remains consistently high. The shift in which metric is most affected—sensitivity at low prevalence vs. specificity at high prevalence—reflects the differing roles of false positives and false negatives under different base rates, emphasizing the importance of contextualizing evaluation metrics to disease prevalence.